\documentclass[12pt]{iopart}
\usepackage{graphicx}
\usepackage{xcolor}

\renewcommand{\vec}[1]{\mbox{\boldmath $#1$}}

\begin{document}

\title
[Two-neutron halo nuclei in one dimension]
{
Two-neutron halo nuclei in one dimension: dineutron correlation and 
breakup reaction}

\author{K. Hagino$^1$, A. Vitturi$^{2,3}$, F. P\'erez-Bernal$^4$, and 
H. Sagawa$^5$}
\address{
$^1$
Department of Physics, Tohoku University, Sendai, 980-8578,  Japan}
\ead{hagino@nucl.phys.tohoku.ac.jp}

\address{$^2$
Dipartimento di Fisica Galileo Galilei, 
Universit\`a di Padova, I-35131 Padova, Italy}
\address{$^3$ INFN, Sezione di Padova, I-35131 Padova, Italy}

\address{$^4$ 
Departamento de Fisica Aplicada, 
Facultad de Ciencias Experimentales, 
Universidad de Huelva, 
21071 Huelva, Spain}

\address{$^5$
Center for Mathematics and Physics, University of Aizu, 
Aizu-Wakamatsu, 965-8580, Fukushima, Japan}

\begin{abstract}
We propose a simple schematic model for two-neutron halo nuclei. 
In this model, the two valence neutrons move 
in a one-dimensional mean field, interacting with each other 
via a density-dependent contact interaction. 
We first investigate the ground state properties, and 
demonstrate that the dineutron correlation can be realized with 
this simple model due to the admixture of even- and odd-parity single-particle 
states. 
We then solve the time-dependent two-particle Schr\"odinger equation 
under the influence of a time-dependent one-body external field, 
in order to 
discuss the effect of dineutron correlation on 
nuclear breakup processes. The time evolution of two-particle 
density shows that the dineutron correlation 
enhances the total breakup probability, especially for the two-neutron 
breakup process, in which both the valence neutrons are promoted 
to continuum scattering states.  
We find that the interaction between the two particles definitely 
favours a spatial correlation of the two outgoing particles, which are mainly emitted in the same direction.
\end{abstract}

\submitto{\JPG}
%\maketitle

\section{Introduction}

Neutron-rich nuclei have attracted much interest during the past 
decades\cite{T95,J04,HJ87,BE91,Zhukov93,JRFG04}, 
and this will continue to be so due to new generation 
radioactive beam facilities in the world. These nuclei 
are characterized by 
a small binding energy, and 
many new features originating from the weakly bound property 
have been found. 
A halo and skin structures with a large spatial extension 
of the density distribution\cite{T85}, 
a narrow momentum distribution \cite{K88}, 
a modification of shell structure and magic numbers\cite{Otsuka01}, 
and strong concentration of 
electric dipole ($E1$) transition
at low excitation energies \cite{F04,A99,N06}, are 
well-known examples. 

Among neutron-rich nuclei, 
two-neutron halo nuclei are particularly intriguing systems 
to study. 
Their structure has 
often been described as a three-body system consisting of two
valence neutrons interacting with each other and with the core
nucleus \cite{BE91,Zhukov93,BH07,HS07}. 
Some light neutron-rich nuclei, such as 
$^{11}$Li and $^6$He, do not have a bound state in the two-body 
subsystem with a valence neutron and a core nucleus. 
These nuclei are referred to as Borromean nuclei, and 
their properties have been studied extensively both 
experimentally \cite{T95,J04,T85,K88,A99,N06} 
and theoretically 
\cite{HJ87,BE91,Zhukov93,JRFG04,EBH97,HS05,HSNS09,MKTI07}. 
Nuclear breakup reactions of Borromean nuclei have also been 
investigated 
 by the continuum-discretized-coupled-channels (CDCC) 
method \cite{MHO04,MEO06,MKY10,RGA08,RGA09} and by the eikonal 
method \cite{OKYS01,BHE98,BCDS09}.

One of the most important current open questions 
concerning the Borromean 
nuclei is to clarify the characteristic nature of correlations between 
the two valence neutrons, which do not form a bound state in the vacuum. 
A strong dineutron correlation, where the two neutrons take 
a spatially compact configuration, has been theoretically 
predicted\cite{HJ87,BE91,Zhukov93,HS05,HSCS07,MMS05,PSS07,M06}. 
Although 
the recent experimental observation of 
the strong low-lying dipole strength distribution in the $^{11}$Li 
nucleus \cite{N06} has provided an 
experimental signature of 
the existence of dineutron correlation in this nucleus, its direct 
evidence has not yet been obtained. 

The pair transfer reaction is another promising way to probe 
the dineutron correlation, as the cross section is known to be 
sensitive to the pairing correlation\cite{vOV01,PBBVB09}. 
However, the reaction dynamics is rather complicated and 
the role of dineutron in the pair transfer reaction 
has not yet been fully clarified, 
although experimental studies on pair transfer reaction with 
exotic nuclei have been initiated recently  \cite{T08,C08,L09}. 
Apparently, it is urged to construct a theoretical framework for pair 
transfer which fully takes into account the pairing correlation and 
the dineutron correlation in its consequence. 

The aim of this paper is to develop a simple toy model for two-neutron 
halo nuclei. Albeit its schematic nature, such model is rather useful as it 
allows detailed studies on the static and dynamical properties of 
two-neutron halo nuclei. 
A schematic model 
is also useful to deepen our understanding of two-neutron halo nuclei 
by providing intuitive pictures of several dynamical processes. 
To this end, we consider a three-body model in which the motion of 
two valence neutrons is confined within one-dimensional spatial space. 
The neutrons are assumed to move in a one-dimensional Woods-Saxon potential, 
while interacting with each other via a two-body interaction, 
which we take a density-dependent contact interaction \cite{BE91,EBH97,HS05}. 
This is a natural extension of the model developed in Ref. \cite{DV09} for a one-neutron 
halo nucleus. We apply the model to the ground state as well as to a 
nuclear breakup 
process of a loosely-bound two-neutron halo nucleus, leaving an application 
to the pair transfer process in a separate publication. 
In passing, a similar one-dimensional three-body 
model has been used in atomic physics, 
in order to discuss {\it e.g.}, the mechanism of the 
double ionization process of He atom by 
intense laser fields \cite{PGB91,GE92,LGE00}. 

The paper is organized as follows. 
In Sec. II, we detail the one-dimensional three-body model. 
In Sec. III, we apply the model to the ground state of a loosely-bound 
two-neutron halo nucleus and discuss the dineutron correlation. 
In Sec. IV, we consider the two-neutron halo nucleus under the influence 
of a time-dependent external field. We solve the time-dependent Schr\"odinger 
equation
 in one-dimensional coordinate space 
 and discuss the time evolution of the two-particle 
wave function. We then summarize the paper in Sec. V. 

\section{One-dimensional three-body model}

We consider a two-neutron halo nucleus in a one dimensional space. 
Denoting the coordinate of the 
two neutrons by $x_1$ and $x_2$, the three-body Hamiltonian reads,
\begin{equation}
H=-\frac{\hbar^2}{2m}\frac{d^2}{dx_1^2}+V(x_1)
-\frac{\hbar^2}{2m}\frac{d^2}{dx_2^2}+V(x_2)
+v_{nn}(x_1,x_2),
\label{H}
\end{equation}
where $m$ is the nucleon mass. Here, we neglect for simplicity 
the recoil kinetic 
energy of the core nucleus. $V(x)$ is the neutron-core potential, 
which we take the Woods-Saxon form\cite{DV09}, 
\begin{equation}
V(x)=-\frac{V_0}{1+e^{(|x|-R)/a}}.
\label{V}
\end{equation}
$v_{nn}$ is the neutron-neutron interaction. We take the 
density-dependent contact interaction for it, that is, 
\begin{equation}
v_{nn}(x,x')=-g\left(1-\frac{1}{1+e^{(|x|-R)/a}}\right)\delta(x-x'),
\end{equation}
where we assume that the density is given by the same functional form 
as the mean-field potential, Eq. (\ref{V}). 
This is the so called surface type pairing interaction, which almost vanishes near 
the center of the core nucleus at 
$x\sim0$. 
The density-dependent contact interaction has been successfully employed 
in the Hartree-Fock-Bogoliubov 
calculations 
for neutron-rich medium-heavy and heavy nuclei\cite{MMS05,BHR03}, as well as 
in describing the structure of $^{11}$Li and $^6$He nuclei with the three-body model 
\cite{BE91,EBH97,HS05}. 

In order to obtain the ground state wave function of the two-neutron halo nucleus, we first solve 
the Schr\"odinger equation for the two-body subsystem, 
\begin{equation}
\left[-\frac{\hbar^2}{2m}\frac{d^2}{dx^2}+V(x)-\epsilon_n\right]\phi_n(x)=0. 
\end{equation}
Each eigenstate $n$ is assumed to have a two-fold degeneracy, depending on the direction of 
the spin of neutron. The continuum states are discretized by putting the nucleus within a large box. 
Denoting the size of the box to be $X_{\rm box}$, we impose the vanishing 
boundary condition $\phi_n(x)=0$ at $x=\pm X_{\rm box}$. 

We expand the two-particle wave function $\Psi(x_1,x_2)$ with the 
single-particle wave 
functions $\phi_n(x)$ as
\begin{equation}
\Psi_{\rm gs}(x_1,x_2)=\sum_{n\leq n'}\alpha_{nn'}\Psi_{nn'}(x_1,x_2), 
\label{expansion}
\end{equation}
where
\begin{eqnarray}
\Psi_{nn'}(x_1,x_2)&=&\frac{1}{\sqrt{2(1+\delta_{n,n'})}}\,
[\phi_n(x_1)\phi_{n'}(x_2)+\phi_n(x_2)\phi_{n'}(x_1)]\,|S=0\rangle.
\label{expansion2}
\end{eqnarray}
Here we have assumed that the spin of the two neutrons form the total spin of $S$=0 configuration, 
so that the wave function is symmetric with respect to the interchange of $x_1$ and $x_2$. 
Notice that the Hamiltonian (\ref{H}) conserves the parity. 
Since the ground state has 
positive parity, 
the expansion in Eq. (\ref{expansion}) 
can therefore 
be restricted to the configurations where the states $n$ and $n'$ have the same parity.  
The expansion coefficients $\alpha_{nn'}$ are determined by diagonalizing the Hamiltonian matrix, whose 
matrix elements read,
\begin{eqnarray}
\langle \Psi_{n_1n_2}|H|\Psi_{n_1'n_2'}\rangle &=& 
(\epsilon_{n_1}+\epsilon_{n_2})\,\delta_{n_1,n_1'}\delta_{n_2,n_2'} \nonumber \\
&&-\frac{4g}{\sqrt{2(1+\delta_{n_1,n_2})}\sqrt{2(1+\delta_{n_1',n_2'})}} \nonumber \\
&&\times \int^\infty_{-\infty}dx
\left(1-\frac{1}{1+e^{(|x|-R)/a}}\right) 
\phi^*_{n_1}(x)\phi^*_{n_2}(x)\phi_{n_1'}(x)\phi_{n_2'}(x).\nonumber \\
\label{melement}
\end{eqnarray}

Once the ground state wave function $\Psi_{\rm gs}(x_1,x_2)$ is obtained, the two-particle and one-particle 
densities can be constructed as 
\begin{equation}
\rho_2(x_1,x_2)=|\Psi_{\rm gs}(x_1,x_2)|^2,
\end{equation}
and
\begin{equation}
\rho_1(x)=\int dx' \rho_2(x,x'), 
\end{equation}
respectively. 
The mean value of the neutron-neutron distance and 
the distance between the core and the center of mass of the two neutrons are also computed as 
\begin{eqnarray}
x_{\rm nn} &=& \sqrt{\int^\infty_{-\infty}dx_1dx_2\,(x_1-x_2)^2\rho_2(x_1,x_2)},\\
x_{\rm c2n} &=& \sqrt{\int^\infty_{-\infty}dx_1dx_2\,\left(\frac{x_1+x_2}{2}\right)^2\rho_2(x_1,x_2)}, 
\end{eqnarray}
respectively. 

\section{Ground state properties} 

We now apply the one-dimensional model to a weakly-bound two-neutron 
halo nucleus. To this end, we take $R=1.27\times 9^{1/3}$ fm, $V_0=-50.085$ 
MeV, and $a$=0.67 fm for the Woods-Saxon potential, Eq. (\ref{V}), 
in order to mimic the $^{11}$Li nucleus. 
This potential possesses four bound 
single-particle 
states. We assume that the lowest three 
bound states are already occupied by the neutrons in the core nucleus, and 
exclude those states explicitly from the expansion in 
Eq. (\ref{expansion}). 
In this way, the Woods-Saxon potential has effectively only one bound state 
with odd parity at $\epsilon_{\rm WS}=-0.15$ MeV. 
That is, the three-body system is bound by 0.3 MeV without the pairing 
interaction. 
Although we could set up a situation in which there is no bound 
state in a Woods-Saxon potential, 
%as in Borromean nuclei, 
we introduce a 
loosely bound single-particle state in order to mock up 
a resonance state, which plays an important role in the $^{11}$Li and $^6$He 
nuclei \cite{HSNS09}. 

\begin{figure}[htb]
\begin{center}
\includegraphics[clip,scale=0.4]{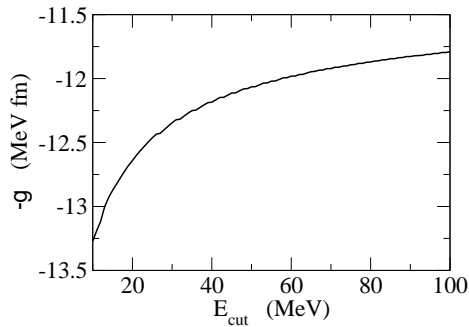}
\caption{
The dependence of the strength of the pairing interaction $g$ 
on the cutoff energy, $E_{\rm cut}$. 
For each $E_{\rm cut}$, the strength $g$ is adjusted in order to 
yield the ground state at $E_{\rm g.s.}=-1$ MeV. }
\end{center}
\end{figure}

It is well known 
that a contact interaction has to be supplemented with an 
energy cutoff, $E_{\rm cut}$ 
\cite{BE91,EBH97}. 
For a three-dimensional case, 
the strength of the pairing interaction, $g$, 
is related to the cutoff energy, $E_{\rm cut}$, 
via a scattering length $a_{\rm nn}$ for the $n+n$ scattering \cite{BE91,EBH97}. 
On the other hand, for a one-dimensional case, 
we simply vary the value of $g$ for each 
cutoff energy 
(with keeping the Woods-Saxon potential well) 
in order to reproduce a given ground state energy, 
since it is not straightforward to define the phase shift for a one-dimensional scattering 
problem. 
Figure 1 shows the dependence of $g$ on 
$E_{\rm cut}$ for the ground state energy of 
$E_{\rm g.s.}=-1$ MeV, obtained by 
taking 
the configurations in Eq. (\ref{expansion}) which satisfy 
$\epsilon_n+\epsilon_{n'}\leq E_{\rm cut}$. 
One notices that the dependence is rather weak. 
In this paper, we arbitrarily 
take the cutoff energy to be 30 MeV,  
that corresponds to $g=-12.35$ MeV$\cdot$ fm for $E_{\rm g.s.}=-1$ MeV. 

\begin{figure}[htb]
\begin{center}
\includegraphics[clip,scale=0.4]{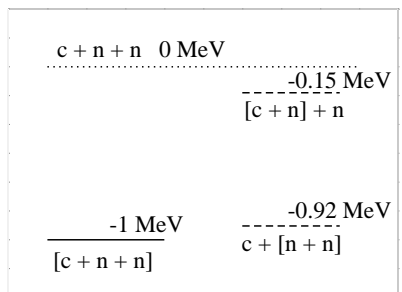}
\caption{
The energy spectrum for the model Hamiltonian 
used in this paper. 
The energies are measured from the threshold 
of a three-body scattering state. 
The three-body bound state, [c+n+n], is located at $E=-1$ MeV. 
The [n + n] and [c + n] are two-body bound states for the two-neutron (n+n) 
and the core+neutron (c+n) systems, respectively. }
\end{center}
\end{figure}

It is important to notice that a one-dimensional delta function potential 
$v(x)=-g\,\delta(x)$ always 
holds a bound state at $E=-mg^2/4\hbar^2$ for a two-neutron 
system  
even with an infinitesimally 
small attraction $g$ \cite{QM-book}. 
For our choice of 
$g=-12.35$ MeV$\cdot$ fm, 
a dineutron is thus bound by 0.92 MeV. 
See Fig.2 for a spectrum for the three-body system considered 
in this paper. There is no bound excited state below the threshold of the 
core nucleus + a bound dineutron at $E=-0.92$ MeV. 
Thus, the ground state in 
the present model has an extremely small separation 
energy from the threshold for the di-neutron breakup, {\it i.e.,} 
less than 100 keV.
Since the dineutron is bound, 
%%%%%%%
our calculations correspond to a limit of 
strong neutron-neutron interaction in a three-dimensional case. 
Alternatively, 
our calculations also 
%%%%%%%%
have a similarity to $^{6}$Li in which a bound deuteron may exist in 
the channel of $^6$Li = $\alpha+p+n$. 

\begin{figure}[htb]
\begin{center}
\includegraphics[clip,scale=0.4]{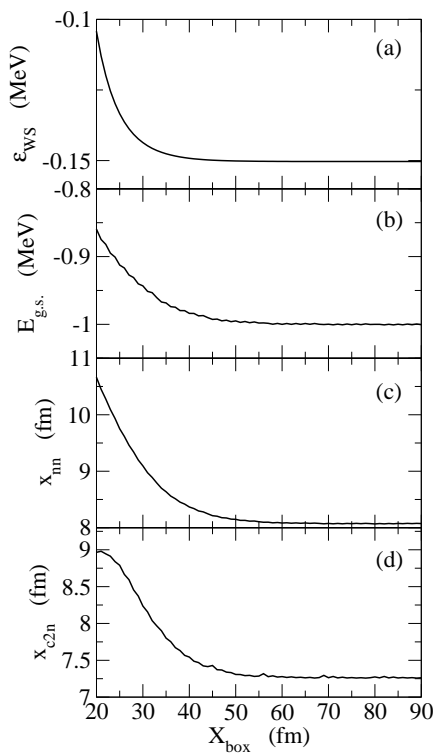}
\caption{
The energy of the 
last  
bound single-particle level (Fig. 3(a)), the ground 
state energy of the three-body system (Fig. 3(b)), the neutron-neutron 
mean distance (Fig. 3(c)), and the core-dineutron mean 
distance (Fig. 3(d)) as a function of the size of the box, $X_{\rm box}$. }
\end{center}
\end{figure}

Figure 3 shows several quantities for the ground state 
of the three-body Hamiltonian, Eq. (\ref{H}), as a function of the size 
of the box, $X_{\rm box}$. 
Figs. 3(a), 3(b), 3(c), and 3(d) are for the 
single-particle energy of the 
last  
bound level, the ground state energy of 
the three-body system, the neutron-neutron root mean square distance, and 
the mean distance between the core and the center of mass of the two neutrons, 
respectively. These quantities are almost converged at $X_{\rm box}\sim$ 50 fm. 
For a Borromean system, it was shown that the convergence of the ground 
state wave function is much slower than the ground state energy, especially 
in the tail region\cite{PBV09,VPB10}. For the example shown in Fig. 3, 
in which there is a bound single-particle state, 
we have 
confirmed that the wave function has also converged 
at $X_{\rm box}\sim$ 50 fm. 
In the calculations shown in this paper, we take the box size to be $X_{\rm box}$=90 fm. 

\begin{figure}[htb]
\begin{center}
\includegraphics[clip,scale=1.2]{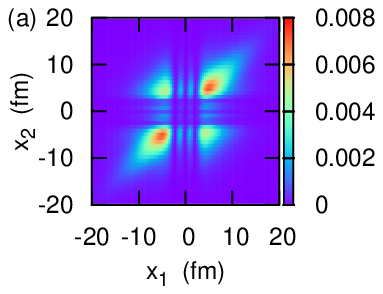}
%\vspace*{-0.7cm}
\includegraphics[clip,scale=1.2]{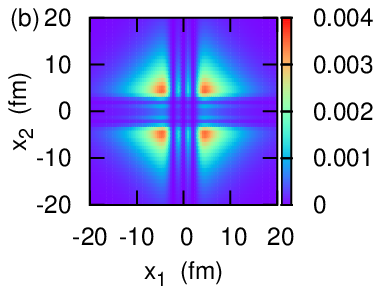}
\caption{(Color online)
The two-particle density for the correlated (Fig. 4(a)) 
and the non-correlated (Fig. 4(b)) 
ground states of a one-dimensional three-body model.}
\end{center}
\end{figure}

Figure 4(a) shows the two-particle density of the ground state. 
The density distribution is largely concentrated along the line 
of $x_1\sim x_2$, 
that is nothing but the manifestation of strong dineutron correlation 
discussed in Refs.\cite{BE91,Zhukov93,HS05,HSCS07,MMS05,PSS07,M06}.
The correlation largely hinders the density 
along the $x_1=-x_2$ line, and only the peaks along $x_1\sim x_2$ survive. 
When the pairing interaction is switched off, the two-particle density 
becomes that shown in Fig. 4(b). The density for the 
non-correlated ground state is symmetric with respect to the transformation 
of $x_1\to -x_1~(x_2\to -x_2)$ for a fixed value of $x_2~(x_1)$, and 
it has therefore four symmetric peaks.
See Ref. \cite{HSS09} for a similar figure for the $^{11}$Li nucleus. 
\begin{figure}[htb]
\begin{center}
\includegraphics[clip,scale=1.2]{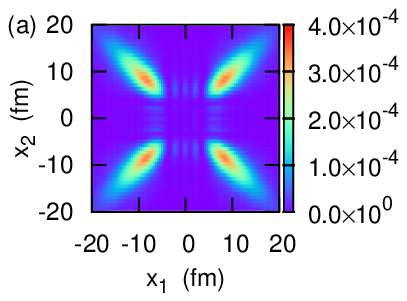}
\hspace*{-1.4cm}
\includegraphics[clip,scale=1.2]{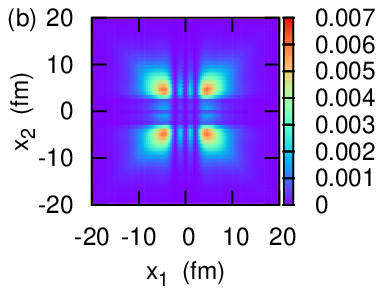}
\hspace*{-1.7cm}
\includegraphics[clip,scale=1.2]{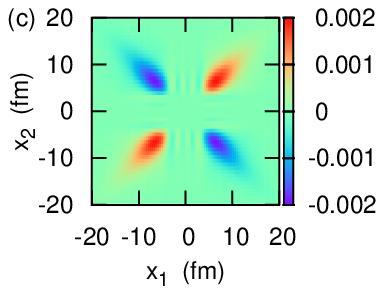}
\caption{(Color online) 
The decomposition of the correlated two-particle density 
into the even-parity contribution (Fig. 5(a)), the odd-parity 
contribution (Fig. 5(b)), and the interference between the even and odd 
parity contributions (Fig. 5(c)). }
\end{center}
\end{figure}

It is well known that the dineutron correlation 
is caused by the admixture of many single-particle states with different 
parity \cite{PSS07,CIMV84}. 
In order to demonstrate it explicitly, we decompose the ground state 
wave function, Eq. (\ref{expansion}), into two components, 
\begin{equation}
\Psi_{\rm gs}(x_1,x_2)=\Psi_{\rm ee}(x_1,x_2)+\Psi_{\rm oo}(x_1,x_2),
\end{equation}
where $\Psi_{\rm ee}(x_1,x_2)$ and $\Psi_{\rm oo}(x_1,x_2)$ 
consist only of the even-parity and the odd-parity 
single-particle states, respectively. 
Assuming that $\Psi_{\rm ee}$ and $\Psi_{\rm oo}$ are both real, 
the two-particle density then reads,
\begin{eqnarray}
\rho_2(x_1,x_2)&=&
|\Psi_{\rm ee}(x_1,x_2)|^2+|\Psi_{\rm oo}(x_1,x_2)|^2 
+2\Psi_{\rm ee}(x_1,x_2)\Psi_{\rm oo}(x_1,x_2). 
\label{eodecomposition}
\end{eqnarray}
Fig. 5 shows each of the components separately 
(see also Fig. 12 in Ref. \cite{PSS07}). 
With only even or odd 
parity single-particle states, 
the density distribution 
has four symmetric peaks as in the 
non-correlated case shown in Fig. 4(b). 
$|\Psi_{\rm oo}|^2$ is much more compact than $|\Psi_{\rm ee}|^2$ because 
the former contains the bound single-particle state. 
It is clearly seen in 
Fig. 5(c) that the localization of the two-neutron pair
 along the line $ x_1= x_2$ 
emerges 
only with the interference 
between the even and odd parity states\cite{PSS07,CIMV84}. 

This can be understood easily as follows. Because the nuclear interaction 
is short ranged, the main contribution comes from $n=n'$ in 
Eq. (\ref{expansion}). 
Suppose that there are only one even-parity single-particle state $\phi_e(x)$ 
and one odd-parity state, $\phi_o(x)$. 
If we write the ground state wave function as 
\begin{equation}
\Psi_{\rm g.s.}(x,x')=\alpha \,\phi_e(x)\phi_e(x')+\beta\, \phi_o(x)\phi_o(x'),
\label{twolevel}
\end{equation}
the two-particle density reads
\begin{eqnarray}
\rho_2(x,x')&=&
\alpha^2\,|\phi_e(x)\phi_e(x')|^2+\beta^2\,|\phi_o(x)\phi_o(x')|^2 \nonumber \\
&&+2\alpha\beta \,\phi_e(x)\phi_e(x')\phi_o(x)\phi_o(x'). 
%%\label{eodecomposition}
\end{eqnarray}
As we prove in the Appendix 
by using the two-level model,
$\alpha$ and $\beta$ have the same sign 
for the density-dependent pairing interaction. 
The first and the second terms in this equation are positive definite. 
On the other hand, the third term is positive for $x=x'$, while it is negative 
for $x=-x'$ as $\phi_e(-x)=\phi_e(x)$ and $\phi_o(-x)=-\phi_o(x)$. 
Therefore, the interference term  is 
 destructive and decreases  the two-body density along the $x=-x'$ line, 
while it is constructive and enhances the density  along the $x=x'$ line.

In the past, the dineutron correlation was discussed using several 
representations. In Refs. \cite{BE91,MMS05,HSS09}, the density distribution 
of the second neutron was investigated when the first neutron was fixed at 
a certain position. In Refs. 
\cite{Zhukov93,PSS07,HSCS07,CIMV84,Dasso,Liotta,Bort,TTD98}, 
the two-particle 
density was plotted as a function of the relative distance between 
the neutrons, $\vec{r}=\vec{r}_1-\vec{r}_2$, and the center of mass 
coordinate,  $\vec{R}=(\vec{r}_1+\vec{r}_2)/2$. 
In Ref. \cite{HS05}, by setting $r_1=r_2=r$ 
the two-particle density was plotted 
as a function of 
the core-neutron distance, $r$, 
and the opening angle between the two neutrons, $\theta_{12}$.  
Although all the representations 
are useful to reveal the strong dineutron 
correlation, they have advantages and disadvantages.  
For instance, it is not easy to explore 
all the position of the first neutron in the first representation, and 
the configurations with $r_1\neq r_2$ are neglected in the third representation. 
These can be avoided with the second representation, but it is more 
intuitive to use directly the coordinates $\vec{r}_1$ and $\vec{r}_2$, 
rather than $\vec{r}$ and $\vec{R}$, especially when 
one has to consider also the angular part of  
$\vec{r}$ and $\vec{R}$. 
The one-dimensional model removes these inconveniences, yielding 
a transparent picture for the two-particle density distribution.

\section{Nuclear Breakup process} 

Let us next discuss the time evolution of the two-particle wave function 
in the presence of an external field acting on each particle. 
As the external field, we take \cite{DV09}
\begin{equation}
V_{\rm ext}(x_1,x_2,t)=\sum_{i=1,2}
V_c\,e^{-t^2/2\sigma_t^2}e^{-(x_i-x_0)^2/2\sigma_x^2}, 
\label{Vext}
\end{equation}
with the parameters of $V_c$=3 MeV, $\sigma_t=2.1 \hbar$/MeV, and 
$\sigma_x$=2.2 fm. In order to investigate the time evolution, 
we solve the time-dependent two-particle Schr\"odinger equation, 
\begin{equation}
i\hbar\frac{\partial}{\partial t}\Psi(x_1,x_2,t)=
[H+V_{\rm ext}(x_1,x_2,t)]\Psi(x_1,x_2,t),
\end{equation}
with the initial condition of 
\begin{equation}
\Psi(x_1,x_2,t_0)=\Psi_{\rm gs}(x_1,x_2),
\end{equation}
at the initial time $t_0$. 
We 
%orthogonalize 
orthonormalize 
the wave function to the occupied states using the 
projection procedure at each time step during the time evolution. 
In the 
calculations shown below, we take $ct_0=-400$ fm, and use the implicit 
method \cite{KDM77} for the time propagation. 
We consider the symmetric perturbation, that is, $x_0=0$ 
in Eq. (\ref{Vext}). We have confirmed that our conclusions remain 
qualitatively the same (except for the asymmetry in the two-particle 
density distribution along the $x_1=x_2$ line) 
even if we choose an asymmetric perturbation \cite{DV09}, {\it e.g.,} with 
$x_0=2$ fm. 
 
\begin{figure*}[htb]
\hspace*{-1.5cm}
\vspace*{-0.7cm}
\includegraphics[clip,scale=1.]{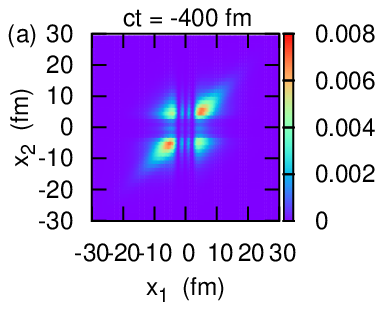}
\hspace*{-1.5cm}
\includegraphics[clip,scale=1.]{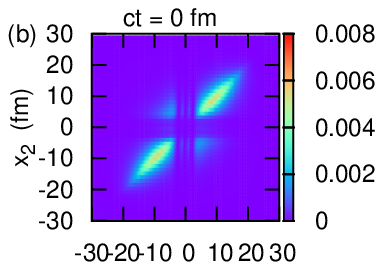}
\hspace*{-1.5cm}
\includegraphics[clip,scale=1.]{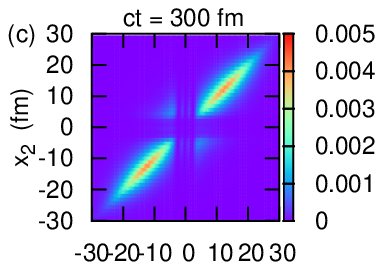}
\hspace*{-1.5cm}
\includegraphics[clip,scale=1.]{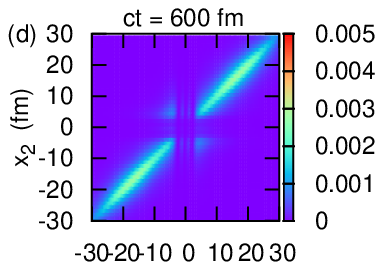} \\
\hspace*{2.6cm}
\vspace*{-0.8cm}
\includegraphics[clip,scale=1.03]{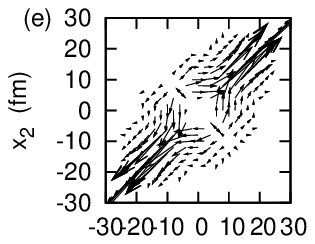}
\hspace*{-0.3cm}
\includegraphics[clip,scale=1.03]{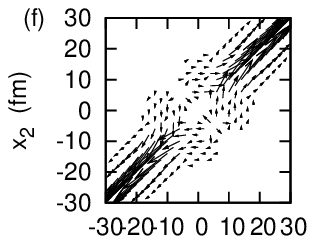}
\hspace*{-0.3cm}
\includegraphics[clip,scale=1.03]{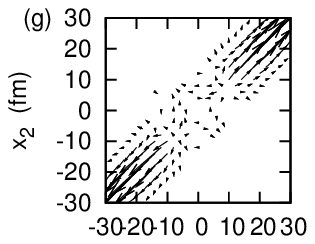} \\
%\vspace*{-1.2cm}
%\hspace*{3cm}
\hspace*{2.2cm}
\includegraphics[clip,scale=1]{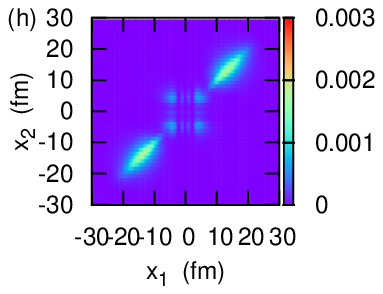}
\hspace*{-1.5cm}
\includegraphics[clip,scale=1]{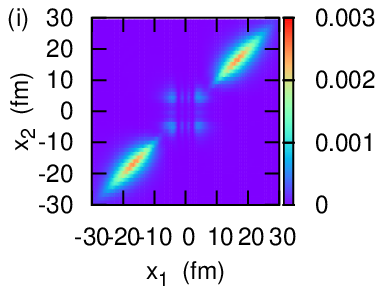}
\hspace*{-1.5cm}
\includegraphics[clip,scale=1]{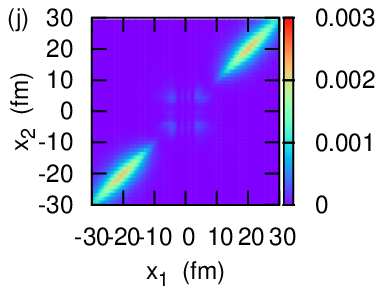} 
\caption{(Color online)
The time evolution of the two-particle density. 
Figs. 6(a), 6(b), 6(c), and 6(d) show the two-particle density, 
$|\Psi(x_1,x_2)|^2$ at $ct=-400, 0, 300,$ and 600 fm, respectively. 
Figs. 6(e), 6(f), and 6(g) show the corresponding current 
distributions with arbitrary units. 
Figs. 6(h), 6(i), and 6(j) show the breakup component of 
the two-particle density, 
$|\Psi_{\rm bu}(x_1,x_2)|^2$ at $ct=0, 300,$ and 600 fm, respectively. 
Notice the difference scales among Figs. 6(a), 6(b), 6(c), and 6(d).  
}
\end{figure*}

\begin{figure*}[htb]
\hspace*{-1.5cm}
\vspace*{-0.7cm}
\includegraphics[clip,scale=1.]{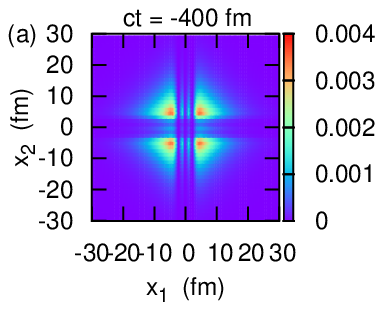}
\hspace*{-1.5cm}
\includegraphics[clip,scale=1.]{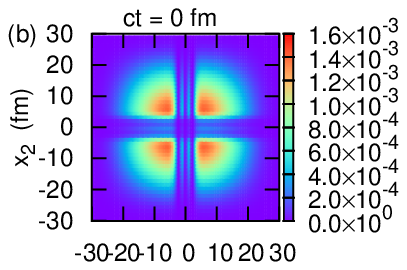}
\hspace*{-1.5cm}
\includegraphics[clip,scale=1.]{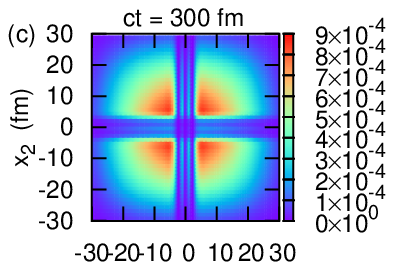}
\hspace*{-1.5cm}
\includegraphics[clip,scale=1.]{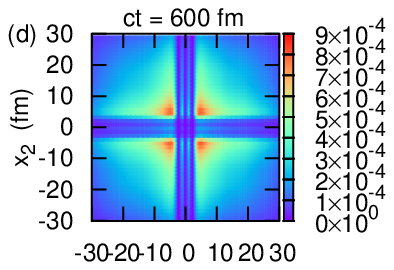} \\
\hspace*{2.6cm}
\vspace*{-0.8cm}
\includegraphics[clip,scale=1.03]{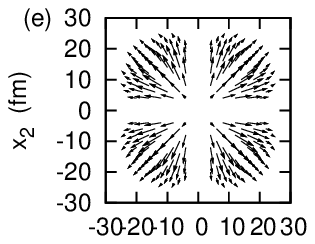}
\hspace*{-0.3cm}
\includegraphics[clip,scale=1.03]{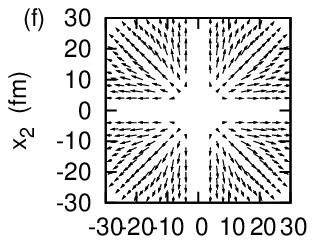}
\hspace*{-0.3cm}
\includegraphics[clip,scale=1.03]{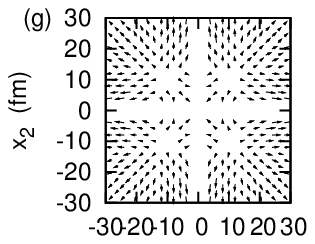} \\
%\vspace*{-1.2cm}
%\hspace*{3cm}
\hspace*{2.2cm}
\includegraphics[clip,scale=1]{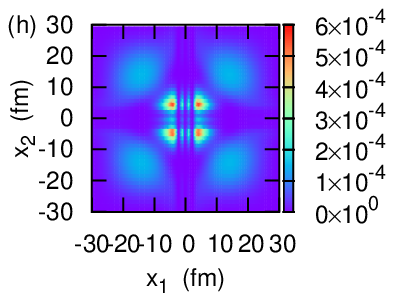}
\hspace*{-1.5cm}
\includegraphics[clip,scale=1]{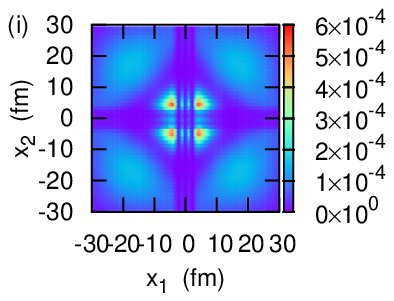}
\hspace*{-1.5cm}
\includegraphics[clip,scale=1]{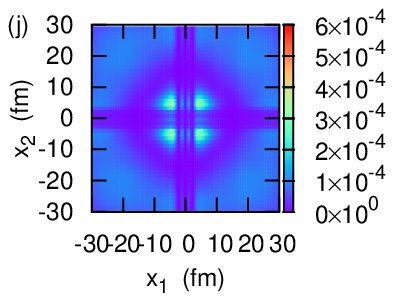} 
\caption{(Color online) 
Same as Fig. 6, but for the uncorrelated case with a vanishing 
neutron-neutron interaction. 
Notice the difference scales among Figs. 7(a), 7(b), 7(c), and 7(d). 
}
\end{figure*}

Figures 6(a), 6(b), 6(c), and 6(d) show the two-particle density 
at $ct=-400, 0, 300$, and 600 fm, respectively. 
As the time evolves, the extension of the peaks along the $x_1=x_2$ line 
increases significantly. This is in marked contrast to the uncorrelated case
shown in Figs. 7(a), 7(b), 7(c), and 7(d). 
In the uncorrelated case, the two-particle density expands 
democratically, indicating that there is the equal probability 
of emission of 
the two neutrons in the opposite directions to that in the same direction. 

The corresponding current distributions, 
\begin{eqnarray}
j_i(x_1,x_2,t)&=&\frac{\hbar}{2im}\left(
\Psi^*(x_1,x_2,t)\frac{\partial}{\partial x_i}
\Psi(x_1,x_2,t)\right. \nonumber \\
&&-\left.
\Psi(x_1,x_2,t)\frac{\partial}{\partial x_i}
\Psi^*(x_1,x_2,t)\right),
\end{eqnarray}
are shown in Figs. 6(e-g) and 7(e-g) for the correlated and the uncorrelated 
cases, respectively. 
For the correlated case, the main flows are the outgoing flows 
along the $x_1=x_2$ line, as one can infer from Figs. 6(b), 6(c), and 6(d). 
For the uncorrelated case, on the other hand, there are four symmetric 
outgoing flows that correspond to the expanding density distribution shown 
in Figs. 7(b), 7(c), and 7(d). 

\begin{figure}[htb]
\begin{center}
\includegraphics[clip,scale=1.2]{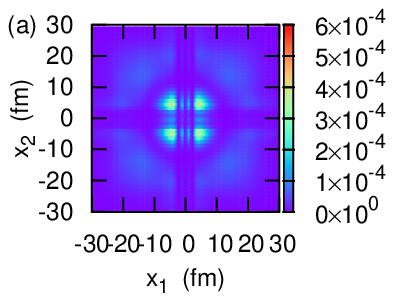}
\includegraphics[clip,scale=1.2]{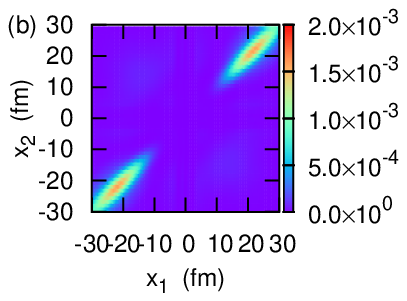}
\caption{(Color online)
The decomposition of the breakup component of 
two-particle density at $ct=600$ fm shown in Fig. 6(j). 
Fig. 8(a) shows the (bc) component, in which one of 
the neutrons is in a continuum state while the other remains in 
the bound single-particle state. Fig. 8(b) shows the (cc) 
component, in which both the two neutrons are in continuum 
states.}
\end{center}
\end{figure}

In order to see more clearly the time evolution of the breakup 
fragments, 
we project the two-particle wave function at each time $t$ onto the 
component which is orthogonal 
to the ground state. 
That is, 
\begin{equation}
\Psi_{\rm bu}(x_1,x_2,t)=\Psi(x_1,x_2,t)-
\langle \Psi_{\rm gs}|\Psi(t)\rangle\,
\Psi_{\rm gs}(x_1,x_2). 
\label{eq:wf-bu}
\end{equation}
Since there is only one three-body bound state (the ground state)
below the dineutron threshold, 
all the remaining wave functions (\ref{eq:wf-bu}) 
necessarily correspond to the breakup processes.
Figs. 6(h), 6(i), and 6(j) show the time evolution of the 
breakup component of the two-particle density, 
$|\Psi_{\rm bu}(x_1,x_2,t)|^2$. 
One can clearly see that 
the two neutrons fly apart from the core nucleus by sticking to each other 
due to the final state interaction. The effect of the correlation is clearly evidenced by the comparison with the uncorrelated case, shown in  Figs. 
7(h), 7(i), and 7(j). 

In order to clarify the dynamics of breakup process, 
we further decompose the breakup component 
of the two-particle wave function into three contributions,
\begin{eqnarray}
\Psi_{\rm bu}(x_1,x_2,t)&=&\Psi^{\rm (bu)}
_{\rm bb}(x_1,x_2,t)+
\Psi^{\rm (bu)}_{\rm bc}(x_1,x_2,t)
+\Psi^{\rm (bu)}_{\rm cc}(x_1,x_2,t),
\label{expansion3}
\end{eqnarray}
where $\Psi_{\rm bb}$ and $\Psi_{\rm cc}$ 
correspond to the component in which both 
the neutrons are in the bound single-particle state and in the continuum 
states, respectively. 
$\Psi_{\rm bc}$ describes the component in which one of the neutrons is 
in the bound state while the other is in a continuum state. 
That is, 
\begin{eqnarray}
\Psi^{\rm (bu)}_{\rm bb}(x_1,x_2,t)&=&\alpha^{(\rm bu)}(t)\,\Psi_{00}(x_1,x_2), 
\label{expansion_bb}
\\
\Psi^{\rm (bu)}_{\rm bc}(x_1,x_2,t)&=&\sum_{n\neq 0}\beta^{(\rm bu)}_{n}(t)\,\Psi_{0n}(x_1,x_2),
\label{expansion_bc}
 \\
\Psi^{\rm (bu)}_{\rm cc}(x_1,x_2,t)&=&
\sum_{{n\leq n'}\atop {(n,n'\neq0)}}\gamma^{(\rm bu)}_{nn'}(t)\,\Psi_{nn'}(x_1,x_2), 
\label{expansion_cc}
\end{eqnarray}
where $\Psi_{nn'}$ is given by Eq. (\ref{expansion2}), $n=0$ corresponding 
to the bound single-particle level at $-$0.15 MeV.  
Figs. 8(a) and 8(b) show the (bc) and (cc) components of the breakup density, 
$|\Psi_{\rm bc}(x_1,x_2)|^2$ and $|\Psi_{\rm cc}(x_1,x_2)|^2$, respectively, 
at $ct=600$ fm shown in Fig. 6(j). 
The (bb) component is small, and is not shown in the figure 
(the norm of 
$\Psi_{\rm bb}$, $\Psi_{\rm bc}$, and $\Psi_{\rm cc}$ are 
4.58$\times 10^{-4}$, 0.116, and 0.466, respectively. See also 
Fig. 11 below). 
It is 
interesting to notice that 
the inner part of the 
density for the (bc) component somewhat 
resembles the non-correlated density at early times 
shown in Fig. 7(h). 
As we will show in Fig. 10, this originates from the fact that 
the (bc) component makes a dominant contribution in the uncorrelated case.  
For the (cc) component, on the other hand, 
the two emitted neutrons are sticking to each other due to the final 
state interaction, and the breakup process 
may be interpreted as 
an emission of a bound dineutron. 

\begin{figure}[htb]
\begin{center}
\includegraphics[clip,scale=1.2]{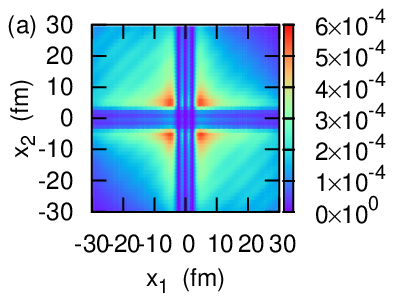} 
\includegraphics[clip,scale=1.2]{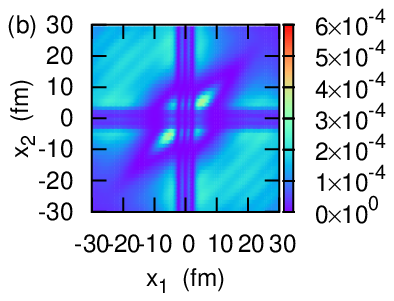}\\
\includegraphics[clip,scale=1.2]{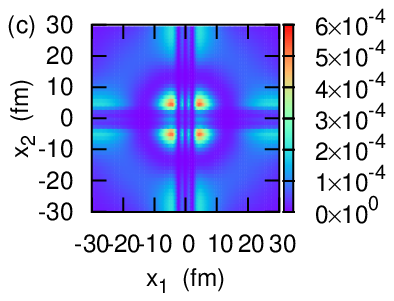} 
\includegraphics[clip,scale=1.2]{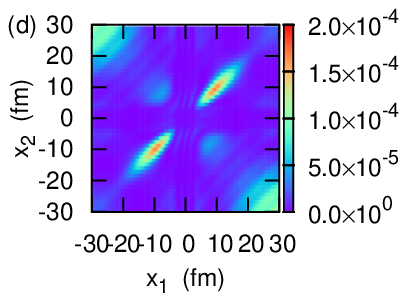}
\caption{(Color online) 
The two-particle densities at $ct=600$ fm obtained by 
neglecting the final state interaction ({\it i.e.,} 
the neutron-neutron interaction) during the time evolution 
from the correlated ground state at initial time. 
Figs. 9(a) and 9(b) show the total density, $|\Psi(x_1,x_2)|^2$, and its 
breakup component, $|\Psi_{\rm bu}(x_1,x_2)|^2$, respectively, while 
Figs. 9(c) and 9(d) show 
the (bc) and the 
(cc) components of the breakup density, respectively. }
\end{center}
\end{figure}

We point out that both the dineutron correlation in the ground state, 
%and the final state interaction during the time evolution are 
as well as 
the neutron-neutron interaction acting during the time evolution 
(that is, the final state interaction) 
are 
important for the dineutron emission process. 
In order to demonstrate this, Fig. 9 shows the two-particle density 
at $ct=600$ fm obtained by switching off the neutron-neutron interaction 
during the time evolution. 
The initial state is still prepared as the correlated ground state. 
Figs. 9(a) and 9(b) show 
the total density, $|\Psi(x_1,x_2)|^2$, and its 
breakup component, $|\Psi_{\rm bu}(x_1,x_2)|^2$, respectively. The (bc) and the 
(cc) components of the breakup density are also shown in Figs. 9(c) and 
9(d), respectively. 
One can see that 
the density distributions are considerably different from those corresponding 
to the fully correlated case, Figs. 6(d), 6(j), 8(a), and 8(b). Without the 
final state interaction, the probability for the non-correlated two-neutron 
emission is increased significantly (see the peaks in Figs. 9(a), 
9(b) and 9(d) 
along the $x_1=-x_2$ line). At the same time, the dineutron emission 
is somewhat slowed down, as the peaks in Figs. 9(d)  
along the $x_1=x_2$ line are located 
much closer to the core nucleus as compared to the fully correlated case 
shown in Fig. 8(b). 
%Evidently, the final state interaction is essential 
%in the breakup processes. 

\begin{figure}[htb]
\begin{center}
\includegraphics[clip,scale=1.2]{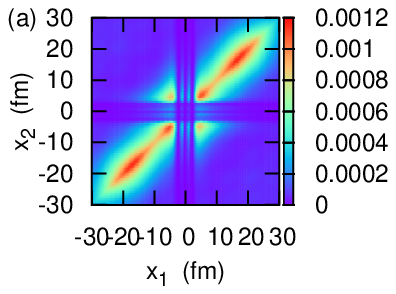} 
\includegraphics[clip,scale=1.2]{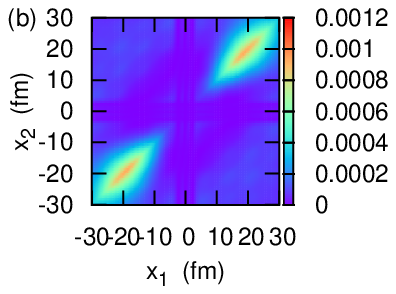}
\caption{
(Color online) 
Same as Figs. 9(a) and 9(b), but for the case 
with the final state interaction during the time evolution 
from the {\it uncorrelated} ground state at initial time. } 
\end{center}
\end{figure}

Figure 10(a) and 10(b) show the calculated results
starting with the {\it uncorrelated} initial state, but with 
taking into account 
the neutron-neutron interaction during the time evolution. 
To this end, we multiply the same time profile function, $e^{-t^2/2\sigma_t^2}$, 
as in the time-depenent external field, Eq. \ref{Vext}, to the 
pairing interaction in the time-dependent Schr\"odinger equation. 
Figs. 10(a) and 10(b) show the 
total density, $|\Psi(x_1,x_2)|^2$, and its 
breakup component, $|\Psi_{\rm bu}(x_1,x_2)|^2$, respectively. 
The density along the diagonal $x_1=x_2$ line now increases 
significantly, as in the full correlated 
calculations shown in Figs. 
6(d) and 6(j). 
Evidently, the final state interaction is essential 
in the breakup processes. 

\begin{figure}[htb]
\begin{center}
\includegraphics[clip,scale=0.4]{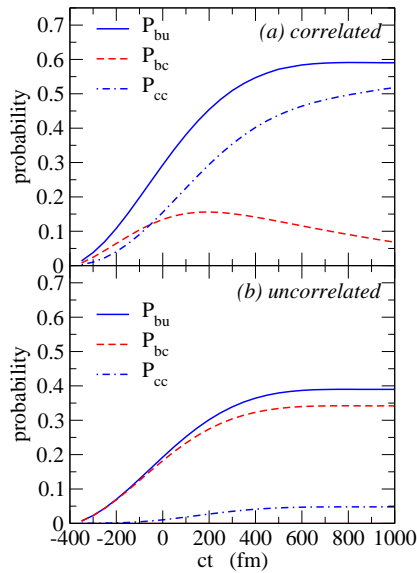}
\caption{(Color online) 
The breakup probability as a function of time. 
The upper and the lower
panels correspond to the correlated and the uncorrelated cases,
respectively.
The solid line shows 
the total breakup probability, while 
the dashed and the dot-dashed lines correspond to decompositions of 
the breakup probability into the (bc) and (cc) processes, respectively.
See the text for details. 
}
\end{center}
\end{figure}

The total breakup probability as well as the probability for each 
breakup process are shown in Fig. 10 as a function of time. 
They are defined as 
\begin{eqnarray}
P_{\rm bu}(t)&=&\int_{-\infty}^{\infty}dx_1dx_2\,|\Psi_{\rm bu}(x_1,x_2,t)|^2, \\
&=&|\alpha^{(\rm bu)}(t)|^2+\sum_{n\neq 0}|\beta^{(\rm bu)}_{n}(t)|^2 
+\sum_{{n\leq n'}\atop {(n,n'\neq0)}}|\gamma^{(\rm bu)}_{nn'}(t)|^2, \\
P_{\rm bc}(t)&=&
\int_{-\infty}^{\infty}dx_1dx_2\,|\Psi^{\rm (bu)}_{\rm bc}(x_1,x_2,t)|^2
=\sum_{n\neq 0}|\beta^{(\rm bu)}_{n}(t)|^2, 
\end{eqnarray}
and
\begin{equation}
P_{\rm cc}(t)=\int_{-\infty}^{\infty}dx_1dx_2\,
|\Psi^{\rm (bu)}_{\rm cc}(x_1,x_2,t)|^2
=\sum_{{n\leq n'}\atop {(n,n'\neq0)}}|\gamma^{(\rm bu)}_{nn'}(t)|^2, 
\end{equation}
which are denoted by the solid, the dashed, and the dot-dashed 
lines, respectively. 
The upper and the lower panels correspond to the correlated and the 
uncorrelated cases, respectively. 
Notice $P_{\rm bb}=0$ for the uncorrelated case because of the 
orthogonalization of the wave functions. 
For the correlated case, 
although $P_{\rm bb}$ is finite, it is considerably small and is 
not shown in Fig. 10. 
By comparing the upper and the lower panels of Fig. 10, one 
finds that the total breakup probability is increased due to the pairing 
correlation. 
This is the case especially for $P_{\rm cc}$, that is, for 
the two-neutron emission process. 

One also 
sees that the (bc) component first increases as a function of time while 
the increase of the (cc) component is somewhat delayed. For the correlated 
case, the (bc) component eventually decreases and the (cc) component takes 
over. This is a manifestation of the dominance of a sequential mechanism in the two-neutron breakup, if 
one intends to use a terminology of perturbation theory. 
Nevertheless, the strong final state interaction makes the dineutron-like 
emission the main breakup process as shown in Fig. 6. 

In Ref. \cite{HSNS09}, we have pointed out that the properties of the 
two-body subsystem with a neutron and the core nucleus play a decisive 
role in the Coulomb breakup of $^{11}$Li and $^6$He nuclei. 
This is because the external field is so weak for the Coulomb breakup that 
only one of the neutrons makes a transition to other (continuum) 
single-particle states. This corresponds to the (bc) process in our 
example. The external field is much stronger for the nuclear breakup 
process, and the two-step process, or even higher step processes, play 
an important role. 
This can be seen in a large probability for the (cc) process, 
$P_{\rm cc}$, 
shown in the upper panel of Fig. 10. 
The effect of dineutron correlation can therefore 
be much easily seen in the nuclear 
breakup process as compared to the Coulomb breakup. 
A similar conclusion has been reached also in Ref. \cite{AL09} 
(see also Ref. \cite{A09}). 

\section{summary}

We have developed a simple schematic model for two-neutron halo nuclei, 
which still contains the essential 
features of physics of unstable nuclei. The model is based on a three-body 
model in one spatial dimension. 
That is, the two valence neutrons move in a one dimensional mean field 
potential while interacting with 
each other via a two-body interaction. 
The two-body interaction scatters the two neutrons into many single-particle 
states with opposite parity, 
causing the strong dineutron correlation in the ground state. 
Applying this model to a weakly 
bound two-neutron halo nucleus, we 
have shown that the two-particle density for the ground state is indeed 
concentrated 
in the region of $x_1\sim x_2$, 
as a consequence of the dineutron correlation. This model also 
allows detailed studies on several dynamical processes. We have solved 
the time-dependent two-particle 
Schr\"odinger equation in a non-perturbed way under the influence of a 
time-dependent external field, which can simulate the field generated by the reaction partner during a heavy-ion collision. 
We have shown that the main breakup process is an emission of dineutron, 
that is, the correlated neutron pair, in spite of the one-body nature of the external field and
although the one neutron emission process is dominant at the early stage 
of time evolution. 
We have also shown that the pairing correlation, and thus the dineutron 
correlation, significantly enhances 
the breakup 
probability, especially for the two-neutron emission process. 

A schematic model such as that presented in this paper is useful to get 
deep insight into the physics behind. 
In addition to the nuclear breakup process studied in this paper, 
the one dimensional model for two-neutron halo nuclei 
can also be used {\it e.g.,} in order to clarify 
the pair transfer reactions of exotic nuclei, whose dynamics has 
not yet been fully understood. 
A work in this direction is now in progress, and we will publish it in a 
separate paper \cite{HV10}. 
Subbarrier fusion reaction of two-neutron halo nuclei is another 
interesting application of the 
one dimensional model. It would be straightforward to extend the one 
dimensional model for fusion of one-neutron 
nuclei \cite{YS95} to two-neutron nuclei. Such study will shed light on 
the effect of irreversible process 
such as breakup on many-particle quantum tunneling. 

\ack

K.H. thanks INFN, Sezione di Padova, 
for its hospitality and financial support.  
This work was supported 
by the Grant-in-Aid for Scientific Research (C), Contract No. 
22540262 and 20540277 from the Japan Society for the Promotion of Science. 

\appendix

\section{Admixture of even- and odd-parity states}
In this appendix, we prove by using a two-level model 
that $\alpha$ and $\beta$ in Eq. (\ref{twolevel})
  have the same sign for  
   the density-dependent pairing interaction. 
These coefficients are determined by diagonalizing a 2$\times$2 matrix, 
\begin{equation}
\left(
\begin{array}{cc}
A&B\\
B&C
\end{array}
\right),
\end{equation}
with
\begin{eqnarray}
A&=&2\epsilon_e-\int^\infty_{-\infty}dx\,g(x)\,\phi_e(x)^4, \\
B&=&-\int^\infty_{-\infty}dx\,g(x)\,\phi_e(x)^2\phi_o(x)^2, \\
C&=&2\epsilon_o-\int^\infty_{-\infty}dx\,g(x)\,\phi_o(x)^4, \\
g(x)&\equiv&g\left(1-\frac{1}{1+e^{(|x|-R)/a}}\right),
\end{eqnarray}
(see Eq. (\ref{melement})). The lower eigenvalue of this matrix is, 
\begin{equation}
\lambda=\frac{1}{2}\,\left[(A+C)-\sqrt{(A-C)^2+4B^2}\right], 
\end{equation}
and the corresponding eigenvector is 
\begin{equation}
\left(
\begin{array}{c}
\alpha \\
\beta
\end{array}
\right)
=
{\cal N}\cdot 
\left(
\begin{array}{c}
2B \\
(C-A)-\sqrt{(C-A)^2+4B^2}
\end{array}
\right),
\end{equation}
where 
${\cal N}$ is the normalization coefficient. 
Since both $B$ and $(C-A)-\sqrt{(C-A)^2+4B^2}$ are negative, 
$\alpha$ and $\beta$ have the same sign, as in the BCS ground 
state wave function in which 
all the configurations are superposed coherently.

\end{document}